# Detecting and Tracking The Real-time Hot Topics: A Study on Computational Neuroscience

Xianwen Wang* and Zhichao Fang

WISE Lab, Faculty of Humanities and Social Sciences, Dalian University of Technology, Dalian 116085, China.

* Corresponding author.
Email address: xianwenwang@dlut.edu.cn; xwang.dlut@gmail.com
Website: http://xianwenwang.com

**Abstract:** In this study, following the idea of our previous paper (Wang, et al., 2013a), we improve the method to detect and track hot topics in a specific field by using the real-time article usage data. With the "usage count" data provided by Web of Science, we take the field of computational neuroscience as an example to make analysis. About 10 thousand articles in the field of Computational Neuroscience are queried in Web of Science, when the records, including the usage count data of each paper, have been harvested and updated weekly from October 19, 2015 to March 21, 2016. The hot topics are defined by the most frequently used keywords aggregated from the articles. The analysis reveals that hot topics in Computational Neuroscience are related to the key technologies, like "fmri", "eeg", "erp", etc. Furthermore, using the weekly updated data, we track the dynamical changes of the topics. The characteristic of immediacy of usage data makes it possible to track the "heat" of hot topics timely and dynamically.

**Keywords**: article usage, usage count, hot topic, topic detection, computational neuroscience

## Introduction

In our previous study (Wang, et al., 2013a), we proposed the method to detect and track the real-time research trends in scientometrics field based on the article downloads data. In this paper, we make further improvement to the methodology and detect and track hot topics in a specific field with the "usage count" data provided by Web of Science (WoS).

For researchers, hot topics are the most attractive research questions. They reflect the scientific research trends that get most attention from researchers, so hot topics in themselves are significant indicators to the strength of attention received from scientific communities. The congregate attention often indicates the major problems to be addressed or the promising new issues in each domain. As a result, it is hoped that hot topics could be used to identify the research interest, emerging trends and major projects for researchers and policy makers of science and technology. Especially for policy makers, monitoring research trends helps resource allocation and technology forecast (Tseng et al., 2009), so detecting and tracking hot topics has far more profound implications.

Hot topics change as time goes by due to the progress of the fields they belong to, particularly when the paradigm shifts happened. The methods for detecting hot topics have been developed a lot during the past decades. At first, detecting hot topics and identifying research trends are regarded as the specialties and tasks of domain experts. Experts are consulted for this work based on their extensive expertise and experience. Even to this day, expert consultation method is still widely used in the process of science and technology policy making. This traditional method plays an incomparable role over time. However, in the age of knowledge explosion, much more information than ever before need to be taken into account. In this case, partial or inconsistent conclusions may be drawn if simply rely on qualitative analyses by different experts. Therefore, quantitative evaluation is introduced into detecting hot topics and tracking upward research trends. Quantitative metrics with high expectations could reveal hot topics effectively and timely and support qualitative, expert assessment (Hicks et al., 2015).

Hot topics may be consisted of words or phrases, which come from the bibliographic data of articles, including article title, abstract, keywords, and so on. These data are deemed to contain a reasonably detailed picture of the article's subject (Garfield, 1990), and are always used to quantitatively analyze research trends or detect topics (Wen & Huang, 2012; Dong et al., 2012; Chen et al., 2015; Tan et al., 2014). But the specific selection methods for detecting hot topics vary with data processing. Traditionally, there are two main methods were used to extract terms as hot topics. One is the statistical analysis or co-occurrence analysis on the publication data. Word frequency analysis (Carroll & Roeloffs, 1969; Zhong & Song, 2008; Su et al., 2014), co-word analysis (Callon et al., 1991; Liu et al., 2012; Gan & Wang, 2015), word cluster analysis (Mao et al., 2010; Zheng et al., 2015), etc. are widely applied to topic detection. And the other is combined with the citation data of scientific articles to identify hotspots, such as co-citation (Chen, 2006; Ding, 2011; Yan & Ding, 2012; Xie, 2015), bibliographic coupling (Glänzel & Czerwon, 1996), h-index (Banks, 2006; Ye, 2013), etc. Nevertheless, previous researches have examined the validity of these methods, both of them are flawed (Healey et al., 1986). The former only takes authors' subjective statements into consideration but ignores the evaluation and reflection made by scientific community. Besides, it cannot handle synonym and polysemy terms very well which presents difficulties in subsequent clustering tasks (Ding & Chen, 2014). The latter has been widely criticized due to the time delay of citations (Amat, 2007; Peng & Zhu, 2012). The existence of time delay makes it difficult to track hot topics timely. Therefore, identifying hot topics only based on bibliographic data and citation data has much limitations.

Compared to bibliographic data and citation data, usage data are a kind of newly developed dataset with unique characteristics. Relatively complete usage behaviors, just like viewing, downloading, clicking links to publishers, are recorded in usage data. Nowadays, more and more academic services are opening these data to public, like PLoS, Nature, Science, Springer, Frontiers, etc. Through usage data, researchers could evaluate researches (Davis et al., 2008), assess or predict impact (Kaplan & Nelson, 2000; Brody et al., 2006; Shuai et al., 2012; Wang et al., 2014a), identify usage patterns or rules (Wang et al., 2014b), and explore users' behaviors (Davis and Solla 2003; Davis and Price 2006; Wang et al. 2012, 2013b). In addition, usage behaviors show the interest that users have. The more attractive the topic of an article is, the higher the usage count it will get. So usage data are supposed to be used to predict new research trends (Wang et al. 2013a). Compared with bibliographic data and citation data, usage data could reflect the evaluations to an article that users make, and even more timely and sensitive. Usage data provide a new way to detect and track hot topics.

In this study, we attempt to detect and track the hot topics in the field of computational neuroscience on the basis of article usage data harvested from Web of Science platform. Our research questions are, what keywords are most used within this field? Is it possible to detect and track hot topics timely by usage data?

## Data and Methods

Our research starts with a search for all articles published within the field of computational neuroscience indexed by Web of Science Core Collection. As an interdisciplinary science, computational neuroscience aims at explaining how electrical and chemical signals are used in the brain to represent and process information (Sejnowski et al., 1988). The development of computational neuroscience could have significant impacts on intelligent science, information science, cognitive science, and neurosciences, etc. Furthermore, in our data observation period, from October 19, 2015 to March 21, 2016, the number of articles about computational neuroscience increased from 96,330 to 101,940, this discipline retains a remarkably strong pulse so that the alternation of hot topics could be more measurable and clearer.

And the usage data of these articles are also collected from Web of Science directly. On September 26, 2015, Web of Science released 5.19 version and added item level usage metrics on the Web of Science platform, which called "Usage Count" (Wang et al, 2016). As Figure 1 shows, located on the summary page as well as the full record, a usage count will be displayed for the Last 180 Days (rolling) as well as Since 2013 (all time). The count reflects the number of times the article has met a user's information needs as demonstrated by clicking links to a full-length article at the publisher's website (via direct link or Open URL) or by saving the metadata for later use (http://wokinfo.com/media/pdf/wos_release_519.pdf). Now, we could know how many times an article in Web of Science has been used. The availability of usage data on the Web of Science platform provides a brand new and all-around data source for us to track the hot topics in different fields.

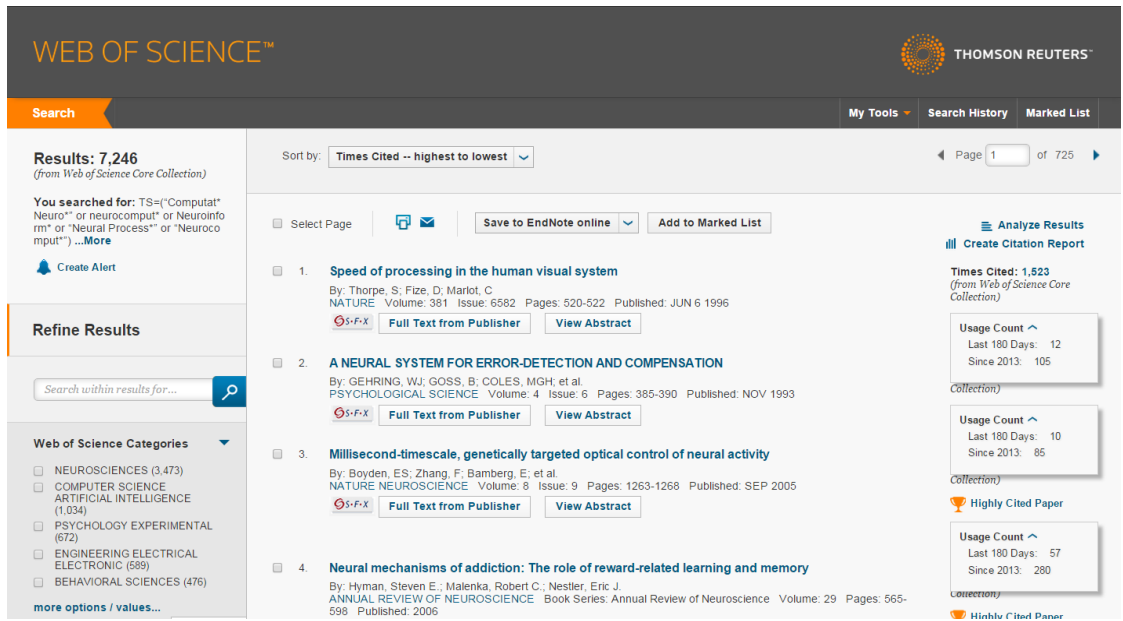

Figure 1 "Usage Count" is displayed on the Web of Science platform

Our research starts with a search, including keywords search and citing articles search, for all articles published within the field of computational neuroscience indexed by WoS. And the specific usage data ("Since 2013" usage count) of these articles were collected weekly from October 19, 2015 to March 21, 2016, along with some main bibliographic data of articles.

The usage of keywords is used to detect the hot topics. For the data collected from Web of Science, if the record has author keywords (DE), then DE extracted from the record is used as keyword; for the records without DE field but have ID field (keywords plus, added in Web of Science), ID is regarded as keyword alternatively; And for those records without DE nor ID, we extract keywords by word segmentation of the article titles. The process of extracting keywords is shown in Figure 2. In order to avoid the interferences caused by single & plural form and stop words, we stemmed each word. These stemmed words or phrases condense the subject of articles, which means that they are the research topics in the field of computational neuroscience.

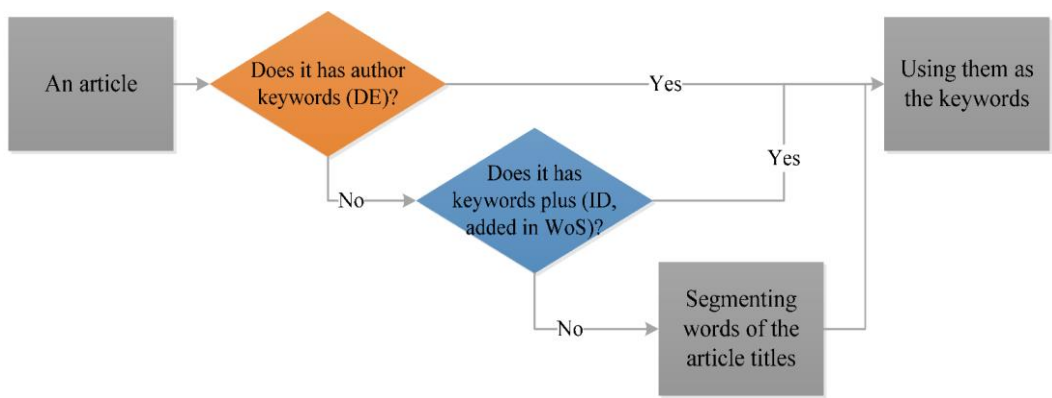

Figure 2 The process of extracting keywords

And the usage of keywords is aggregated from the usage of articles. How many times an article is used indicates how many times its keywords are used, as Figure 3 shows.

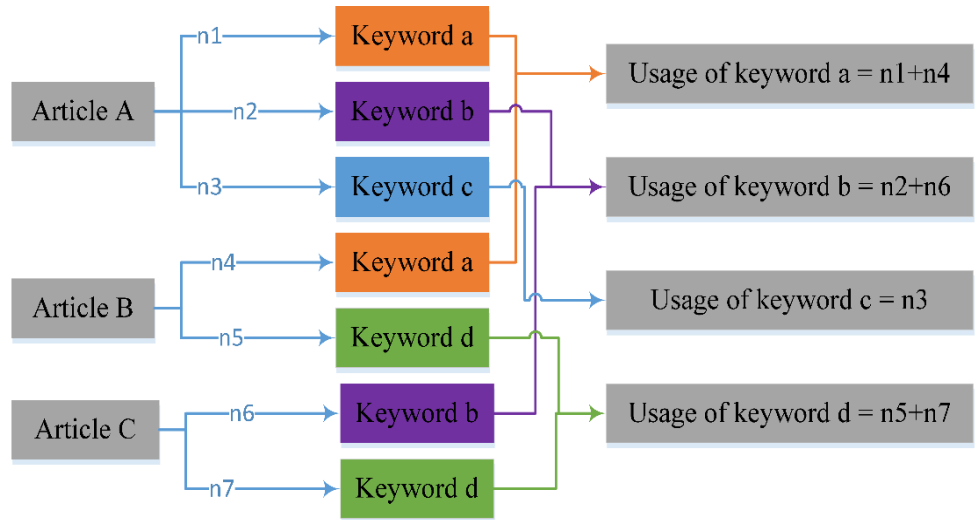

Figure 3 Aggregating article usage to keyword usage

## Results

### Hot topics

After the combination of synonyms, like "fmri", "function mri" and "function magnet reson imag", "erp" and "event-relat potenti", etc. Table 1 lists top 20 topics of most frequent occurrence. They are most used by researchers themselves in writing their articles, which reveal the interest of authors in the field directly. These topics stand out due to the frequent usage in the process of writing by authors without the evaluation by others. They are the hot topics from the perspective of bibliographic data.

Table 1 Top 20 topics with most frequent occurrence

| Rank | Topics | Frequency |
|---|---|---|
| 1 | fmri | 7784 |
| 2 | erp | 4351 |
| 3 | attent | 3535 |
| 4 | prefront cortex | 2849 |
| 5 | memori | 2811 |
| 6 | brain | 2797 |
| 7 | model | 2661 |
| 8 | eeg | 2634 |
| 9 | neural network | 2474 |
| 10 | percept | 2230 |
| 11 | cortex | 2193 |
| 12 | neuron | 2192 |
| 13 | emot | 2190 |
| 14 | hippocampu | 1977 |
| 15 | schizophrenia | 1867 |
| 16 | rat | 1827 |
| 17 | system | 1630 |
| 18 | ag | 1606 |
| 19 | dopamin | 1495 |
| 20 | human | 1474 |

Table 2 lists the top 20 topics with most usage count and their usage count ratio (Ratio1). The articles that these topics located in were frequently used, and that is what these topics brought about. They are always the concern in the field of computational neuroscience, namely the hot topics from the perspective of usage data.

$$\text{Ratio1} = \frac{Total\ usage\ count\ of\ the\ topic}{Total\ usage\ count\ of\ all\ articles}$$

Table 2 Top 20 most used topics

| Rank | Topics | Usage count | Ratio |
|---|---|---|---|
| 1 | fmri | 16652 | 8.07% |
| 2 | erp | 8762 | 4.24% |
| 3 | eeg | 8528 | 4.13% |
| 4 | emot | 6689 | 3.24% |
| 5 | prefront cortex | 6297 | 3.05% |
| 6 | attent | 6011 | 2.91% |
| 7 | memori | 5275 | 2.56% |
| 8 | brain | 5088 | 2.46% |
| 9 | schizophrenia | 4470 | 2.17% |
| 10 | function connect | 3993 | 1.93% |
| 11 | percept | 3849 | 1.86% |
| 12 | bci | 3824 | 1.85% |
| 13 | cortex | 3557 | 1.72% |
| 14 | ag | 3463 | 1.68% |
| 15 | filter | 3321 | 1.61% |
| 16 | work memori | 3286 | 1.59% |
| 17 | reward | 3280 | 1.59% |
| 18 | amygdala | 3252 | 1.58% |
| 19 | rqnn | 3207 | 1.55% |
| 20 | cognit control | 3207 | 1.55% |

There are two main types of these hot topics. One is basic nouns, such as "emot (emotion)", "attent (attention)", "memori (memory)", "brain", "percept", "filter", "reward", "ag (aging)", "model", "rat (rate)", "system", etc. The other is professional terms, such as "fmri (functional magnetic resonance imaging)", "eeg (electroencephalograph)", "prefront cortex (prefrontal cortex)", "schizophrenia", "event-relat potenti (event-related potential)", "bci (brain-computer interface)", "amygdala", "rqnn (recurrent quantum neural network)", "neural network", "hippocampu", "dopamin (dopamine)", "cognit control (cognitive control)", etc. For detecting hot topics, the latter has more research value.

The situations of above hot topics in word frequency rank and usage count rank are shown in Figure 4. The topics located on the horizontal axis are only ranked in top 20 most used, and those on the vertical axis are only ranked in top 20 most frequent occurrence. The others are both frequently used in writing articles by authors and frequently used (saving or clicking) by scientific community.

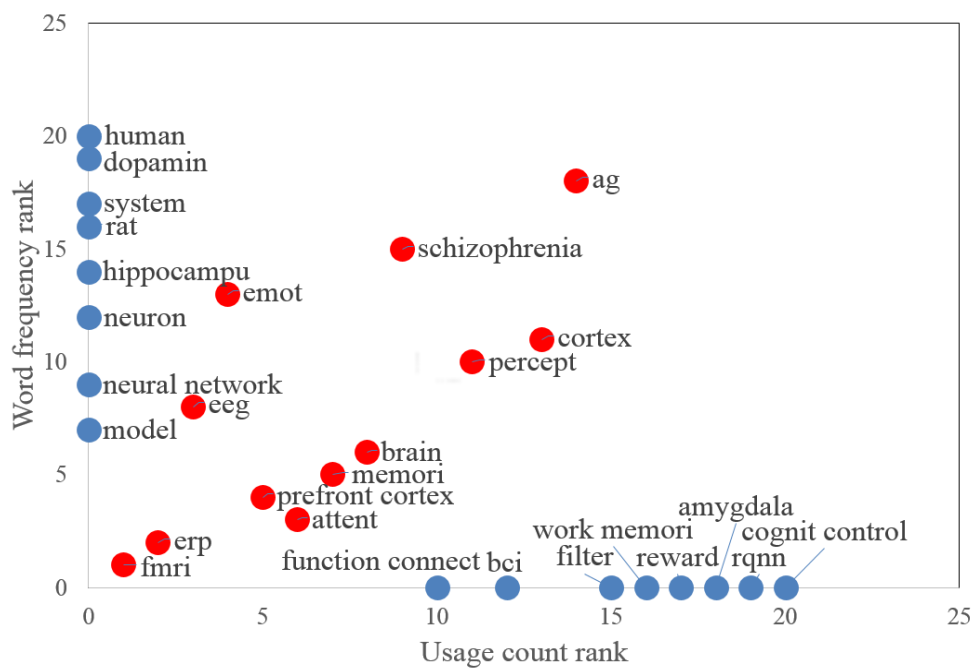

Figure 4 Hot topics in word frequency rank and usage count rank

Overall, some key technologies, like "fmri", "eeg", "erp", were highly used by authors and scientific community, it may result from that these technologies are widely applied to the articles in this field. And usage count analysis identified many different professional terms that word frequency analysis underestimates, like "bci", "rqnn", "amygdala", "cognit control", etc. By contrast, word frequency analysis calculated more basic nouns that have less value in detecting hot topics.

**Dynamical changes of hot topics**

Immediacy is one of the greatest characteristics of usage data. If an article is used, it would be recorded and reflected in short order. It makes it possible to track the hot topics timely. We select some typical hot topics to track their usage count change trends with time goes by. Because the fluctuation of usage count's absolute value is too strong, we calculate their respective percentages (Ratio2) in total usage count of each period. After removing four time periods that the data are incorrect (2016.1.18-2016.1.25, 2016.1.25-2016.2.1, 2016.2.8-

2016.2.15, 2016.3.14-2016.3.21), the time-varying and dynamic "heat" of different hot topics are shown in Figure 5.

$$\text{Ratio2} = \frac{Usage\ count\ of\ the\ topic\ in\ the\ period}{Total\ usage\ count\ of\ all\ articles\ in\ the\ period}$$

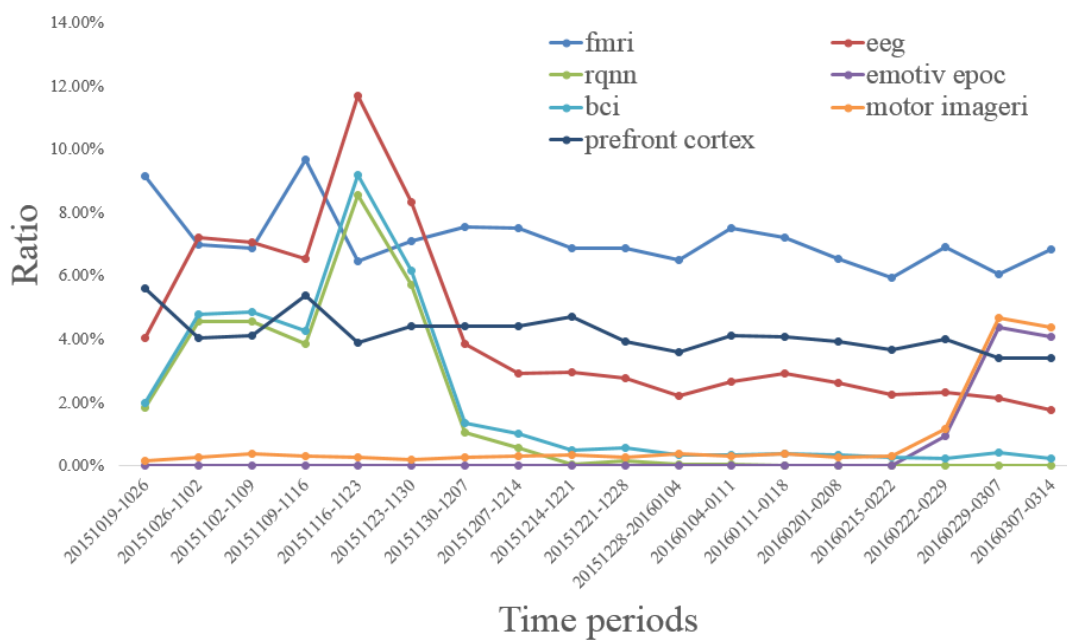

Figure 5 Usage count change trends of typical hot topics

As an important technology, "fmri" maintained high usage count steadily all the while. And "prefront cortex" also maintained relative high usage count steadily as a typical research object. In contrast, "eeg", acts as a general technology too, shows dramatic ups and downs. From November 16, 2015 to November 30, 2015, "eeg" even attained higher usage count than "fmri", but it decreased sharply until February 29, 2016. Same as "eeg", "bci" and "rqnn" show similar trends. They were highly used in its early stage and decreased sharply since November 23, 2015, but unlike "eeg" and "bci", "rqnn" fell into silence and hardly got usage count in the later stage. "emotiv epoc" and "motor imageri" present opposite traces. Until February 22, 2016, both of them were barely used, but from February 22, 2016 to March 14, 2016, they received a lot of attention. "emotiv epoc" and "motor imageri" could be regarded as emerging hot topics.

## Conclusions and discussion

In this study, we detect and track hot topics applying the usage data of scientific articles. Compared to the bibliographic data and citation data used by previous studies, article usage data could reflect the hot topics from another perspective with more detailed and real-time information. The usage count provided by Web of Science opens up a brand new data source. We take computational neuroscience as an example to investigate the feasibility of detecting and tracking hot topics based on usage count. It turned out that many professional terms in this field, such as "fmri", "eeg", "erp", "bci", etc., are identified and regarded as the most used topics. These topics could be defined as the hot topics from the perspective of usage. Compared to word frequency analysis based on bibliographic data, usage data are beneficial to identify some more valuable hotspots. Moreover, through tracking the usage count change trends of different hot topics in each time period, we illustrate that the "heat" of hot topics is time-varying. Some hot topics maintain a stable usage count over time, while some experience a dramatic fluctuation. Therefore, usage data could be used to track the change trends of hot topics timely and dynamically.

There are also some limitations for this study. Firstly, Web of Science defines "usage" as "clicking" and "saving" on the platform of Web of Science, when the number of views and downloads of full texts of the articles are unavailable. However, Web of Science is not the most favored platform for researchers to search, view and download primary research articles, meaning that relative limited usage logs would be accumulated by the platform (Wang et al 2016). Secondly, because Web of Science have disclosed usage data since September 26 2015, our data observation period is a little short to reveal more complete results. We have been collecting the data since October 2015 and improving the methodology, in order to make further and deeper analysis of the real-time usage data to reveal the real-time research trends.

## Acknowledgements

The work was supported by the project of "National Natural Science Foundation of China" (61301227, 71673038), the project of "Growth Plan of Distinguished Young Scholar in Liaoning Province" (WJQ2014009), and the project of "the Fundamental Research Funds for the Central Universities" (DUT15YQ111).

## References

Amat, C. (2007). Editorial and publication delay of papers submitted to 14 selected Food Research journals. Influence of online posting. Scientometrics, 74(3), 379-389.

Banks, M. G. (2006). An extension of the hirsch index: indexing scientific topics and compounds. Scientometrics, 69(1), 161-168.

Brody, T., Harnad, S., & Carr, L. (2006). Earlier web usage statistics as predictors of later citation impact. Journal of the American Society for Information Science & Technology, 57(8), 1060-1072.

Callon, M., Courtial, J. P., & Laville, F. (1991). Co-word analysis as a tool for describing the network of interactions between basic and technological research: the case of polymer chemsitry. Scientometrics, 22(1), 155-205.

Carroll, J. M., & Roeloffs, R. (1969). Computer selection of keywords using word-frequency analysis. American Documentation, 20(3), 227–233.

Chen, C. (2006). Citespace ii: detecting and visualizing emerging trends and transient patterns in scientific literature. Journal of the American Society for Information Science & Technology, 57(3), 359–377.

Chen, H., Jiang, W., Yang, Y., Man, X., & Tang, M. (2015). A bibliometric analysis of waste

management research during the period 1997---2014. Scientometrics, 105(2), 1005-1018.

Davis, P. M., & Price, J. S. (2006). eJournal interface can influence usage statistics: implications for libraries, publishers, and Project COUNTER. Journal of the American Society for Information Science and Technology, 57(9), 1243-1248.

Davis, P. M., & Solla, L. R. (2003). An IP‐level analysis of usage statistics for electronic journals in chemistry: Making inferences about user behavior. Journal of the American Society for Information Science and Technology, 54(11), 1062-1068.

Davis, P. M., Lewenstein, B. V., Simon, D. H., Booth, J. G., & Connolly, M. J. L. (2008). Open access publishing, article downloads, and citations: randomised controlled trial. Bmj, 337(7665), 1720-1720.

Ding, W., & Chen, C. (2014). Dynamic topic detection and tracking: a comparison of hdp, c-word, and cocitation methods. Journal of the Association for Information Science and Technology, 65(10), 2084–2097.

Ding, Y. (2011). Topic-based PageRank on author co-citation networks. Journal of American Society for Information Science and Technology, 62(3), 187–203.

Dong, B., Xu, G., Luo, X., Cai, Y., & Gao, W. (2012). A bibliometric analysis of solar power research from 1991 to 2010. Scientometrics, 93(3), 1101-1117.

Gan, C., & Wang, W. (2015). Research characteristics and status on social media in china: a bibliometric and co-word analysis. Scientometrics, 105(2), 1167-1182.

Garfield, E. (1990). Keywords plus – ISI's breakthrough retrieval method. 1. Expanding Your Searching Power on Current Contents on Diskette. Current Contents, 32, 5-9.

Glänzel, W., & Czerwon, H. J. (1996). A new methodological approach to bibliographic coupling and its application to the national, regional and institutional level. Scientometrics, 37(2), 195-221.

Healey, P., Rothman, H., & Hoch, P.K. (1986). An experiment in science mapping for research planning. Research Policy, 15(5), 233–251.

Hicks, D., Wouters, P., Waltman, L., de Rijcke, S., & Rafols, I. (2015). The Leiden Manifesto for research metrics. Nature, 520(7548), 429-431.

Kaplan, N. R., & Nelson, M. L. (2000). Determining the publication impact of a digital library. Journal of the American Society for Information Science, 51(4), 324-339.

Liu, G. Y., Hu, J. M., & Wang, H. L. (2012). A co-word analysis of digital library field in china. Scientometrics, 91(1), 203-217.

Mao, N., Wang, M., & Ho, Y. (2010). A bibliometric study of the trend in articles related to risk assessment published in Science Citation Index. Human and Ecological Risk Assessment, 16(4), 801–824.

Peng, T. Q., & Zhu, J. J. (2012). Where you publish matters most: A multilevel analysis of factors affecting citations of internet studies. Journal of the American Society for Information Science and Technology, 63(9), 1789-1803.

Sejnowski, T. J., Koch, C., & Churchland, P. S. (1988). Computational neuroscience. Science, 241(4871), 1299-1306.

Shuai, X., Pepe, A., & Bollen, J. (2012). How the scientific community reacts to newly submitted preprints: article downloads, twitter mentions, and citations. Plos One, 7(11), e47523.

Su, X., Deng, S., & Shen, S. (2014). The design and application value of the chinese social science citation index. Scientometrics, 98(3), 1567-1582.

Tan, J., Fu, H. Z., & Ho, Y. S. (2014). A bibliometric analysis of research on proteomics in science citation index expanded. Scientometrics, 98(2), 1473-1490.

Tseng, Y. H., Lin, Y. I., Lee, Y. Y., Hung, W. C., & Lee, C. H. (2009). A comparison of methods for detecting hot topics. Scientometrics, 81(1), 73-90.

Wang, X., Mao, W., Xu, S., & Zhang, C. (2014a). Usage history of scientific literature: nature metrics and metrics of nature publications. Scientometrics, 98(3), 1923-1933.

Wang, X., Peng, L., Zhang, C., Xu, S., Wang, Z., Wang, C., & Wang, X. (2013b). Exploring scientists' working timetable: A global survey. Journal of Informetrics, 7(3), 665-675.

Wang, X., Wang, Z., & Xu, S. (2013a). Tracing scientist's research trends realtimely. Scientometrics, 95(2), 717-729.

Wang, X., Wang, Z., Mao, W., & Liu, C. (2014b). How far does scientific community look back?. Journal of Informetrics, 8(3), 562-568.

Wang, X., Xu, S., Peng, L., Wang, Z., Wang, C., Zhang, C., & Wang, X. (2012). Exploring scientists' working timetable: Do scientists often work overtime?. Journal of Informetrics, 6(4), 655-660.

Wang, X., Fang, Z., & Sun, X. (2016). Usage patterns of scholarly articles on Web of Science: a study on Web of Science usage count.Scientometrics, 1-10. doi:10.1007/s11192-016-2093-0

Wen, H., & Huang, Y. (2012). Trends and performance of oxidative stress research from 1991 to 2010. Scientometrics, 91(1), 51-63.

Xie, P. (2015). Study of international anticancer research trends via co-word and document co-citation visualization analysis. Scientometrics, 105(1), 611-622.

Yan, E.J., & Ding, Y. (2012). Scholarly network similarities: How bibliographic coupling networks, citation networks, cocitation networks, topical networks, coauthorship networks, and coword networks relate to each other. Journal of the American Society for Information Science and Technology, 63(7), 1313–1326.

Ye, F. Y. (2013). Measuring hot topics in sciences. Current Science, 104(2), 160-160.

Zheng, T., Wang, J., Wang, Q., Nie, C., Smale, N., & Shi, Z., et al. (2015). A bibliometric analysis of industrial wastewater research: current trends and future prospects. Scientometrics, 105(2), 863-882.

Zhong, Q., & Song, J. (2008). The Developing Trend Research of Knowledge Management Overseas Based on Word Frequency Analysis. Wireless Communications, Networking and Mobile Computing, 2008. WiCOM '08. 4th International Conference on (pp.1-4). IEEE.